\documentclass[amsfonts,twocolumn,showpacs,pre,aps]{revtex4}
\usepackage{graphicx}
\begin{document}
\title{Nonequilibrium molecular dynamics of complex fluids near the gel point}

\author{Daniel C. Vernon}
\affiliation{Physics Department, Simon Fraser University,
          Burnaby, British Columbia, Canada V5A 1S6}
\affiliation{Department of Physics and Astronomy, University of Pennsylvania,
          Philadelphia, Pennsylvania 19104, USA}
\author{Michael Plischke}
\affiliation{Physics Department, Simon Fraser University,
          Burnaby, British Columbia, Canada V5A 1S6}
\date{\today}
\begin{abstract}
We have carried out nonequilibrium molecular dynamics simulations of a
system of crosslinked particles under shear flow conditions. As the
fraction of crosslinks $p$ is increased the system approaches a gel
point at which the shear viscosity $\eta$ and the first and second
normal stress coefficients $\Psi_1$ and $\Psi_2$ diverge.  All three
quantities seem to diverge with a power law form: $\eta\sim
\epsilon^{-s}$, $\Psi_{1,2}\sim\epsilon^{-\lambda}$ where $\epsilon
=p_c-p$ and $s\approx 0.7$ and $\lambda\approx 3.15$ in three
dimensions and $s\approx 2.0$ and $\lambda\approx 6.0$ in two
dimensions.
\end{abstract}
\maketitle

%\onecolumn
%\narrowtext
\section{Introduction}\label{sec:intro}
There have been many studies, both experimental and theoretical, of
the rheological properties of complex fluids in the vicinity of the
gel point \cite{adam96}. There is general agreement in the literature
that the transition from the sol to the gel phase, at least in the
case of chemical gels, is continuous and accompanied by the divergence
of the shear viscosity $\eta\sim (p_c-p)^{-s}$ where $p$ characterizes
the degree of crosslinking or condensation and $p_c$ is the critical
point.  However, the experimental and theoretical values of the
exponent $s$ are rather widely distributed and there is continuing
debate concerning the existence of a single universality class for
these transitions. We mention, in passing, that we are excluding the
case of vulcanization from this discussion. There is substantial
evidence, in the form of a Ginzburg criterion \cite{degen77}, that the
vulcanization transition which involves the crosslinking of very long
chains is for practical purposes mean-field-like.

There has been much less theoretical and experimental work on the
normal stress differences. In a non-Newtonian fluid under shear flow,
{\it e.g.}, in a Couette geometry with the flow in the $x$-direction
and velocity gradient in the $z$-direction, the first and second
normal stress differences $N_1=\sigma_{xx}-\sigma_{zz}$ and
$N_2=\sigma_{zz}-\sigma_{yy}$ are both not zero and, in the low
shear-rate limit, proportional to the square of the shear rate
$\dot{\gamma}$.  It is conventional to define the normal stress
coefficients $\Psi_1$, $\Psi_2$ through $N_i=\Psi_i\dot{\gamma}^2$.
One of the purposes of this work is to investigate the critical
behavior of the normal stress coefficients.  These are most easily
calculated directly, {\it i.e.}, by shearing the computational cell
and measuring the normal stresses via a virial formula. While there
exists a Green-Kubo formula~\cite{broderix02} that yields the zero shear-rate
limit of $\Psi_1$, the integral involved presents significant computational
problems close to the gel point: the decay of the relevant correlation
function becomes extremely slow and truncation errors become
unmanageable. For this reason, we have decided to utilize
nonequilibrium molecular dynamics (NEMD) to calculate both the shear
viscosity and the normal stress differences.  We use the same model
for which we have previously calculated the shear
viscosity~\cite{MP01} using the Green-Kubo formalism. Our NEMD results
for the critical exponent $s$ are consistent with these earlier results. The
NEMD simulations indicate that the first normal stress coefficient
diverges at the critical point
$\Psi_1(\dot{\gamma}=0)\sim(p_c-p)^{-\lambda}$ with $\lambda\approx
3.15$, {\it i.e.}, much more strongly than does the shear viscosity.
The errors in $\Psi_2(\dot{\gamma}=0)$ are much larger than those in
$\Psi_1$ and an independent determination of an exponent for its
divergence is not feasible. However, the data are consistent with the
conjecture that both normal stress coefficients diverge in the same
way. We note that the conclusion that $\Psi_1$ diverges more strongly
than $\eta$ has also been arrived at by Broderix {\it et
al.}\cite{broderix02} in the context of a Rouse-type model. However,
their exponent $\lambda\approx 4.9$ is significantly larger than
ours.

 The structure of this article is as follows. In section
\ref{sec:model} we briefly describe the model that we have used and
the computational techniques. Section~\ref{sec:dat2} contains our
results for the two-dimensional case and the results in three
dimensions are presented in Section~\ref{sec:dat3}.  We conclude in
Section~\ref{sec:disc} with a brief discussion.

\section{Model and Computational Methods}\label{sec:model}

Our model of the sol phase is identical to that in \cite{MP01}. All
particles interact through the soft sphere potential
$V(r_{ij})=\epsilon(\sigma/r_{ij})^{36}$ for $r_{ij}\leq 1.5\sigma$
and, in the three-dimensional calculations, we have used a single
volume fraction $\Phi=\pi N\sigma^3/6V = 0.4$ which is well below the
liquid-solid coexistence density. All calculations were carried out at
a temperature $k_BT/\epsilon = 1$.  In the absence of crosslinks, this
system is a simple liquid that has been well characterized
\cite{powles}.  We initially placed the particles on the vertices of a
simple cubic lattice and instantaneously and randomly introduced a
fraction $p$ of nearest-neighbor bonds. We used the bonding potential
$V_b(r_{ij})=k(r_{ij}-r_0)^2$ where $k=5\epsilon/\sigma^2$ and where
$r_0=(\pi\Phi)^{1/3}/\sigma$ so that there was no internal mechanical
strain.  This method of crosslinking ensures that the cluster size
distribution is that of percolation in three dimensions and that a gel
forms (in the thermodynamic limit $N\to\infty$) at $p_c\approx
0.2488$.  Once the particles had been crosslinked they were free to
move throughout the three-dimensional computational box. They were
initially thermalized with periodic boundary conditions. Once
equilibrium had been attained, the computational box was sheared at a
rate $\dot{\gamma}=\partial v_x(z)/\partial z$ and the boundary
conditions were changed to the Lees-Edwards boundary conditions
\cite{allen}. The system was then reequilibrated using the so-called
SLLOD algorithm \cite{allen} subject to the constraint that the
kinetic energy in the frame following the overall flow of the
particles remain constant. This kinetic energy is proportional to the
square of the ``peculiar velocity'' $(v_x -\dot\gamma z, v_y, v_z)$.
Once a steady drift had been established, the diagonal elements of the
stress tensor as well as $\sigma_{xz}$ were calculated from the
appropriate virial formula. Calculations were performed for systems of
$N=L^3$ particles with $L$ = 10, 15 and 20 over the entire range
$0\leq p <p_c$ and for shear rates generally in the range
$.005\leq\dot{\gamma}\leq 0.1$ in units of $\sqrt{\epsilon/m\sigma^2}$
although for a few cases larger shear rates were also imposed. It
should be noted that if atomistic values of $\epsilon$, $m$ and
$\sigma$ are used, this range of shear rates corresponds to values of
order $10^{12}s^{-1}$, {\it i.e.}, enormously large rates compared to
experimental values. Even if one takes the point of view that the
particles represent a colloidal suspension, the minimum shear rate is
still of the order of $10^3s^{-1}$. It was necessary to use such large
shear rates in order to obtain reasonably well converged estimates for
the normal stress coefficients, especially close to the gel point. For
this range of shear rates, the equations of motion could be stably
integrated with a time step $\delta t=0.005\sqrt{m\sigma^2/\epsilon}$
but for larger values of $\dot{\gamma}$ the time step had to be
decreased.

The values of the shear viscosity, and to an even larger extent the
normal stress coefficients, varied considerably for different
realizations of the crosslinks. Therefore, we averaged the results
over several thousand realizations even for values of $p$ as small as
0.1.

The same repulsive pair potential was used in the two-dimensional
case.  Here the particles were initially placed on a triangular
lattice and instantaneously crosslinked as in three dimensions. This
sets the gel point at $p_c=2\sin(\pi/18)\approx 0.347296$. The lattice
constant of the initial configuration was $1.2$, and so these
simulations were done at a number density $n\approx 0.8$.  The spring
constant used was $k=40\epsilon/\sigma^2$ and the value of $r_0=1.2$
was chosen to eliminate mechanical strain due to the crosslinks at
zero temperature, as was done in three dimensions.

\section{Results in Two Dimensions}\label{sec:dat2}

\begin{figure}
\includegraphics[width=8.5cm]{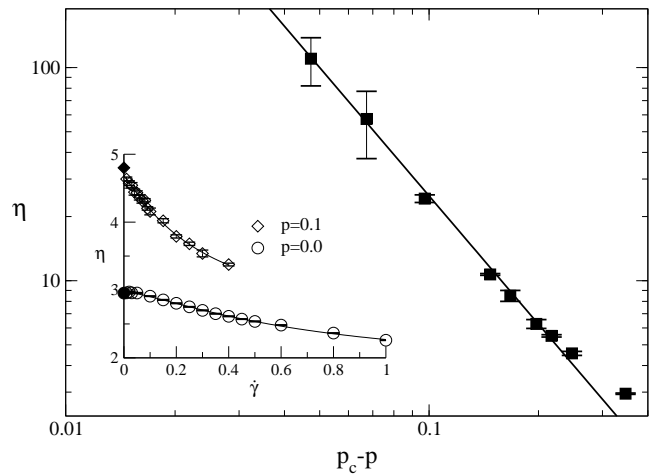}

\caption{The viscosity as a function of $p$ for a crosslinked fluid in
two dimensions.  The straight line shows a power law, with exponent
$s=2$. The inset shows the shear rate dependent viscosity (open
symbols) as compared to the zero shear rate values (filled symbols).
The curves show fits of the finite shear rate data to the Lorentzian
form discussed in the text.}
\label{fig1}
\end{figure}

The final results for the zero shear rate viscosity extrapolated from
nonequilibrium molecular dynamics simulations are shown in
Fig.~\ref{fig1}.  These data are from simulations of systems of size
$32\times32=1024$ particles.  No significant deviation from a scaling
form is visible for $p>0$ at this system size for the range of $p$
studied here.  The power law divergence of the viscosity thus directly
gives an estimate for the exponent $s=2$, without a finite size
scaling analysis.

The inset to Fig.~\ref{fig1} shows the shear rate dependence of the
viscosity as well as the zero shear rate values from the Green-Kubo
formula, for an uncrosslinked and for a lightly crosslinked sample.
The extrapolation of the finite shear rate values to zero shear-rates
seems to be consistent with the Green-Kubo values.  This fluid
exhibits shear thinning, as is seen in some experiments on complex
fluids.  The determination of a zero-shear-rate viscosity requires
fitting to some functional form for the shear-rate dependence of
$\eta$.  However, this value is not sensitive to the form chosen.
Many different functional forms have been suggested for the shear-rate
dependence of the viscosity, mostly suggested as phenomenological
fitting functions to experimental data~\cite{bhw}.  The value for
$\eta$ in the main figure was estimated from a fit to a Lorentzian
plus a constant term, as suggested in~\cite{ferrario92}, while the
difference between different fits was used as an estimate of the
error.  The Lorentzian form has the advantage that it is automatically
symmetric in $\dot\gamma$ and is analytic near $\dot\gamma=0$.

\begin{figure}
\includegraphics[width=8.5cm]{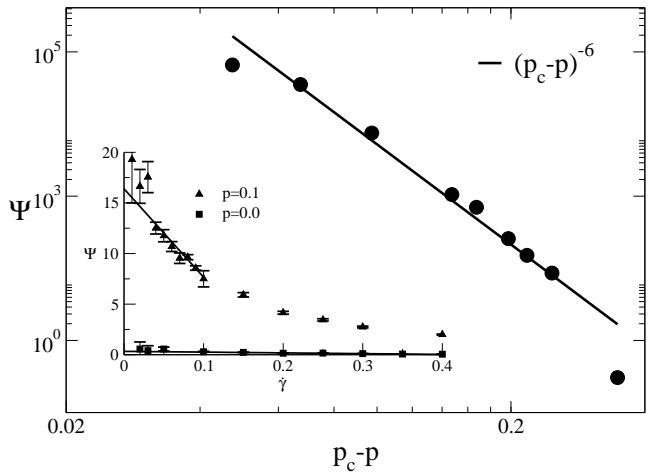}

\caption{The normal stress coefficient $\Psi$ in two dimensions close
to $p_c$.  The straight line is a power law with exponent
$\lambda=6$. The inset shows the linear extrapolation used to
determine the $\dot\gamma=0$ value of $\Psi$.}
\label{fig2}
\end{figure}

There is only one normal stress difference,
$N_1=\sigma_{xx}-\sigma_{yy}$, in a two dimensional system.  We have
measured the associated coefficient $\Psi_1$ close to $p_c$.  This
quantity is expected to diverge as a power law $\Psi_1\sim
(p_c-p)^{-\lambda}$ as the gel transition is approached; the results
of our simulation are shown in Fig.~\ref{fig2}.  We estimate an
exponent $\lambda\approx 6$ from our data.

The estimation of the normal stress coefficient was more difficult
than for the viscosity.  The division by $\dot\gamma^2$ produces large
statistical errors for small $\dot\gamma$, thus making it difficult to
determine the functional form of the shear-rate dependence.  An
example is shown in the inset to Fig.~\ref{fig2}.  Only a simple
linear fit was used to estimate the zero-shear-rate normal stress
coefficient.

\section{Results in Three Dimensions}
\label{sec:dat3}

\begin{figure}
\includegraphics[width=8.5cm]{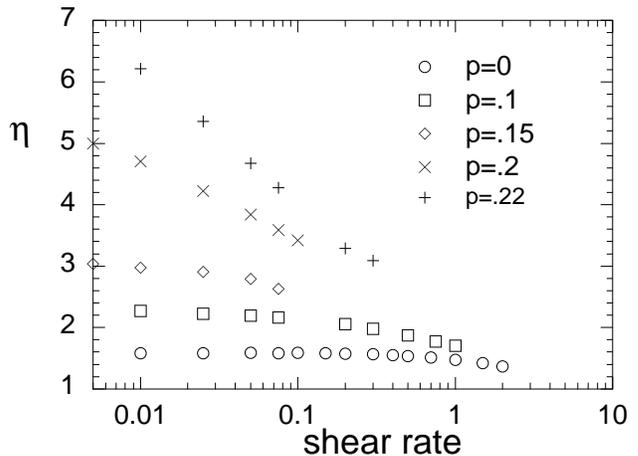}

\caption{Dependence of the shear viscosity $\eta(p,\dot{\gamma})$ on 
the shear rate $\dot{\gamma}$ for $L=10$ and a selection of 
crosslink densities.} \label{fig3}
\end{figure}

\begin{figure}
\includegraphics[width=8.5cm]{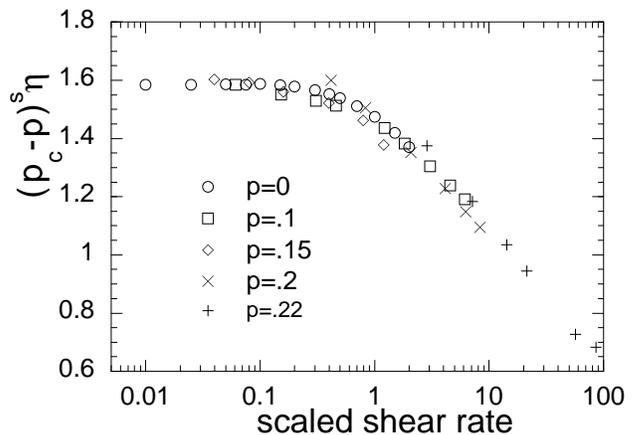}

\caption{The data of figure 1 plotted as function of the scaled shear
  shear rate $\tilde{\dot{\gamma}}=(p_c-p)^{-z}\dot{\gamma}$ with
  $z=2.35$. The  viscosity has been multiplied by $(p_c-p)^s$ with
  $s=0.7$ to remove the divergence at $p_c$.} 
\label{fig4}
\end{figure}

In the three-dimensional case, we have results for systems of $N=L^3$
particles with $L=10$, $15$, and $20$. We first display, in
Fig.~\ref{fig3}, the shear-rate dependence of the viscosity $\eta
(\dot{\gamma})$ for $L=10$ and five different values of the degree of
crosslinking ranging from the simple fluid case, $p=0$, to a system
close to the gel point ($p=0.22$). All systems show evidence of shear
thinning, with this feature becoming much more prominent and setting
in at lower shear rates as the gel point is approached. If we rescale
the shear viscosity using the form $\eta(p,\dot{\gamma})
=a(p_c-p)^s\tilde\eta(p,\tilde{\dot{\gamma}})$ with $s=0.7$ and
$\tilde{\dot{\gamma}}=b(p_c-p)^{-z}\dot\gamma$ where $a$ and $b$ are
constants, we can achieve a respectable collapse of the data, as seen
in Fig.~\ref{fig4}. Since the data are noisy, we have not made a
serious effort to optimize this collapse. Nevertheless, the
``dynamical" exponent $z\approx 2.35$. The exponent $s=0.7$ used to rescale
the viscosity is our best estimate of the exponent that governs the
divergence of the zero shear-rate viscosity
$\eta(p,\dot{\gamma}=0)\sim (p_c-p)^{-s}$. 

We note that dynamical scaling yields a connection between the exponent $z$
that controls the divergence of the longest relaxation time at the gel point
and the exponents $s$ and $t$, {\it i.e.}, $z=s+t$ \cite{adam96}. Here
$t$ is the exponent that describes how the shear modulus vanishes as
the gel point is approached from the solid side: $\mu\sim
(p-p_c)^t$. For this model, we have determined~\cite{mp99} that
$t\approx 2.0$. This yields the prediction $z\approx 2.7$, a value not
too far from the value used to rescale the shear rate.

\begin{figure}
\includegraphics[width=8.5cm]{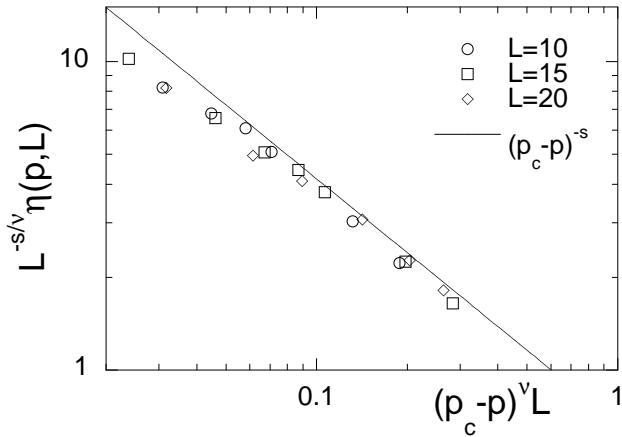}

\caption{Finite size scaling plot of the zero-shear-rate viscosity.}
\label{fig5}
\end{figure}

The zero-shear-rate viscosity is shown in Fig.~\ref{fig5} in finite-size scaled
form, {\it i.e.}, we plot
$L^{-s/\nu}\eta(p,L)$ as function of
$(p_c-p)^\nu L$ where $\nu$ is the correlation length exponent of the three
dimensional percolation problem $\nu\approx 0.88$. As mentioned above,
the value of the exponent $s=0.7$ is consistent with our previous
result obtained from a Green-Kubo calculation \cite{MP01}.

\begin{figure}
\includegraphics[width=8.5cm]{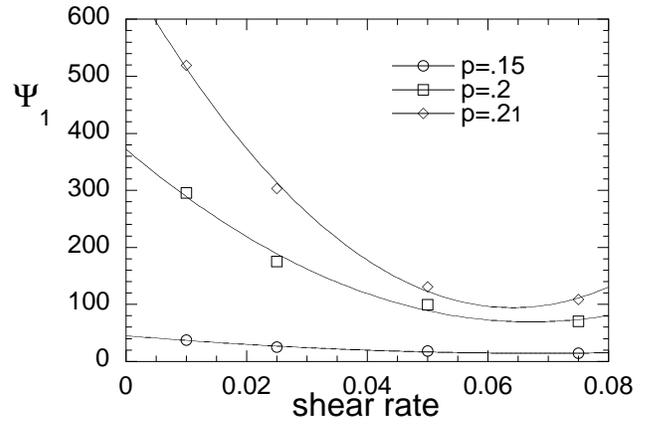}

\caption{First normal stress coefficient for $L=10$ and $p=.15$, $.2$
and $.21$. The curves are fits of the data to a second order
polynomial. Note the rapid increase of $\Psi_1(p,\dot{\gamma}=0)$ as
$p\to p_c$.} 
\label{fig6}
\end{figure}

We turn now to the normal stress coefficients. In Fig.~\ref{fig6} we
plot the first normal stress coefficient
$\Psi_1(p,\dot{\gamma})\equiv(\sigma_{xx}-\sigma_{zz})/\dot{\gamma}^2$
as function of the shear rate for $L=10$ and three values of $p$. For
larger values of $p$, $\Psi_1$ increases rapidly as the shear rate is
decreased and an estimate of the zero shear-rate value
$\Psi_1(p,\dot{\gamma})$ is problematical. We have fit the data points
to a second order polynomial
$\Psi_1(p,\dot{\gamma})=\Psi_1(p,0)+a\dot{\gamma}+b\dot{\gamma}^2$ and
obtained our estimate of the zero shear-rate value in this way. A fit
to an exponential decay works equally well and produces estimates of
$\Psi_1(p,0)$ that differ by no more than 3\% from those shown
here. Similarly, the Lorentzian-plus-constant fit that was used to fit
the viscosity in two dimensions also provides a reasonable fit to the
data as long as there are enough values of the shear rate (more than
five). The conclusions presented below are insensitive to the method
of extrapolation. 

\begin{figure}
\includegraphics[width=8.5cm]{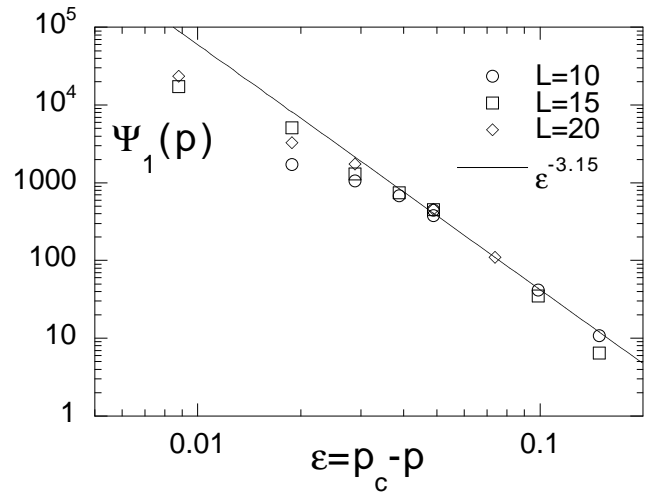}

\caption{The first normal stress coefficient $\Psi_1(p,L)$ plotted as
  a function of
$p_c-p$. The solid line corresponds to a power law divergence of the form
$\Psi_1(p,\infty)\sim (p_c-p)^{-3.15}$.}
\label{fig7}
\end{figure}

\begin{figure}
\includegraphics[width=8.5cm]{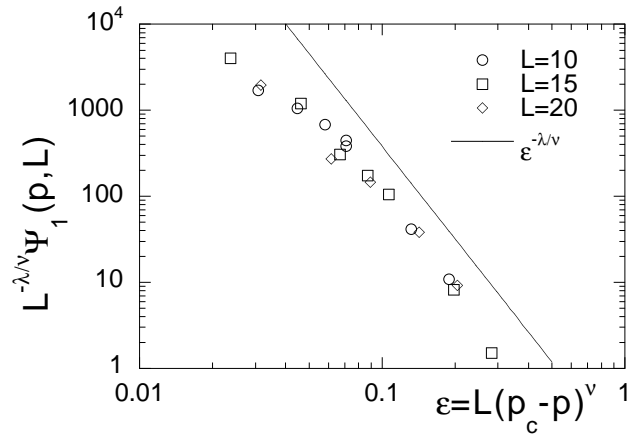}

\caption{The first normal stress coefficient multiplied by $L^{-\lambda/\nu}$ with
$\lambda=3.15$
plotted as a function of $\epsilon=L(p_c-p)^\nu$. The straight line correspond to the
expected large-$\epsilon$ form $\epsilon^{-\lambda/\nu}$.}
\label{fig8}
\end{figure}

The unscaled data for $\Psi_1(p,0)$ are plotted in Fig.~\ref{fig7} for
$L=10$, $15$ and $20$ as function of $p_c-p$ along with a line
representing the function $a(p_c-p)^{-\lambda}$ with $\lambda=3.15$
that captures the form of the data outside the critical region quite
well. Closer to the critical point $p_c\approx 0.2488$, the usual
finite-size effects that appear when the geometric correlation length
$\xi (p)$ approaches the system size, $L$, are evident. These
finite-size effects can be hidden by plotting
$L^{-\lambda/\nu}\Psi_1(p,L)$ as a function of the scaled variable
$\epsilon\equiv L/\xi\propto L(p_c-p)^\nu$ where $\nu\approx 0.88$ is
the correlation length exponent. This is done in Fig.~\ref{fig8} and
the data do collapse reasonably well to a universal curve. A useful
consistency check on this procedure is available far from the critical
point: if the exponents $\lambda$ and $\nu$ are correctly determined
then the scaled normal stress coefficient should approach the
power-law form $\epsilon^{-\lambda/\nu}$ at large $\epsilon$. This
line is also displayed in Fig.~\ref{fig8} and the data are consistent
with the expected behavior.

\begin{figure}
\includegraphics[width=8.5cm]{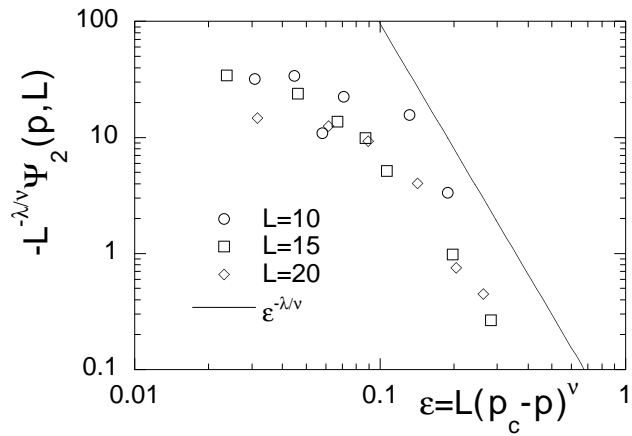}

\caption{Finite size scaling plot of the second normal stress coefficient normal
stress coefficient using the same value of $\lambda$ as in Fig.~\ref{fig8}.}
\label{fig9}
\end{figure}

We have also calculated the second normal stress coefficient
$\Psi_2(p,L)$. This coefficient is much smaller in magnitude than
$\Psi_1$ and negative, at least for our range of shear rates.  The
sample-to-sample fluctuations in $N_2$ are relatively much larger than
those in $N_1$ and the data for $\Psi_2$ are therefore much more
noisy. Indeed, they are so poorly converged that a determination of a
critical exponent from that dataset is not supportable. We have
therefore carried out a rescaling of $\Psi_2$ under the assumption
that it diverges in the same way as $\Psi_1$, {\it i.e.}, controlled
by the same critical exponent $\lambda\approx 3.15$. The results of
this rescaling are shown in Fig.~\ref{fig9} and we see that the data
for $L=15$ and $L=20$ that are somewhat better converged than the data
for $L=10$ seem to support such an assumption.

The authors of reference \cite{broderix02} have proposed a scaling relation
for the exponent $\lambda$: $\lambda= z+s$ that seems to be rigorous for the
Rouse model that they have used. In our case, using the two estimates of $z$
referred to above, namely $z=2.35$ and $z=2.7$, we obtain $\lambda = 3.05$ and
$\lambda= 3.4$ which bracket the measured value. It should however be noted
that in the Rouse model of \cite{broderix02}, the second normal stress
coefficient $\Psi_2=0$ for all values of $p$.

\section{Conclusions}\label{sec:disc}

Using nonequilibrium molecular dynamics simulations, we have measured
the divergence of both the viscosity and the normal stress coefficient
in a model gel as the gel transition is approached.  Our results for
the divergence of the zero shear-rate viscosity are consistent with
our previous calculation using a Green-Kubo formula to extract the
viscosity from an equilibrium simulation.  In addition, this model
exhibits shear thinning as the shear rate is increased, as is observed
in experiments on gelling systems.  We found that the exponent
governing the divergence of the shear viscosity to be $s=0.7$ in three
dimensions.  This value is consistent with some
experiments~\cite{devreux93,durand87},
as well as several analytical calculations~\cite{muller03}.  We also find
$s=2$ in two dimensions.

We have also presented evidence that, in this model, the shear-rate-dependent
viscosity can be rescaled onto a single universal curve. This indicates
that the physics of the shear-thinning that we observe for all $p$ is the
same close to the gel point as it is in the simple liquid ($p=0$).

The divergence of the normal stress close to the gel transition has
not been measured in a experiment.  As suggested in~\cite{broderix02},
it should be possible to observe the very strong divergence of this
quantity experimentally.  Measuring both the viscosity and the normal
stress close to the gel point would give the ratio of two dynamical
exponents without determining the critical point, which is often
difficult to determine accurately in an experiment.  The ratio of $s$
to $\lambda$ would then provide a dynamical exponent which would
characterize the universality class of a given material.  Comparing
this value to the values predicted by different models could then give
some insight into which features of a microscopic model are important
to the dynamical properties of an incipient gel.

\section*{ACKNOWLEDGMENTS}
We thank B\'{e}la Jo\'{o}s for helpful discussions. This research was supported
by the NSERC of Canada, and by the National Science Foundation through
MRSEC grant No.\ DMR~0079909.

\end{document}